\documentclass[a4paper,fleqn,usenatbib]{mnras}

\usepackage{newtxtext,newtxmath}

\usepackage[T1]{fontenc}
\usepackage{ae,aecompl}

\usepackage{graphicx}	
\usepackage{amsmath}	
\usepackage{amssymb}	

\usepackage[normalem]{ulem}
\useunder{\uline}{\ul}{}
\usepackage{bm}

\title[Circularizing Planet Nine]{Circularizing Planet Nine through dynamical friction with an extended, cold planetesimal belt}

\author[L. E. J. Eriksson, A. J. Mustill and A. Johansen]{
Linn E.J. Eriksson,$^{1}$\thanks{E-mail: nat13ler@student.lu.se}
Alexander J. Mustill,$^{1}$
Anders Johansen$^{1}$
\\
$^{1}$Lund Observatory, Department of Astronomy and Theoretical Physics, Lund University, Box 43, SE-221 00 Lund, Sweden\\
}

\date{Accepted XXX. Received YYY; in original form ZZZ}

\pubyear{2017}

\begin{document}
\label{firstpage}
\pagerange{\pageref{firstpage}--\pageref{lastpage}}
\maketitle

\begin{abstract}
Unexpected clustering in the orbital elements of minor bodies beyond the Kuiper belt has led to speculations that our solar system actually hosts nine planets, the eight established plus a hypothetical ``Planet Nine". Several recent studies have shown that a planet with a mass of about 10 Earth masses on a distant eccentric orbit with perihelion far beyond the Kuiper belt could create and maintain this clustering. The evolutionary path resulting in an orbit such as the one suggested for Planet Nine is nevertheless not easily explained. Here we investigate whether a planet scattered away from the giant-planet region could be lifted to an orbit similar to the one suggested for Planet Nine through dynamical friction with a cold, distant planetesimal belt. Recent simulations of planetesimal formation via the streaming instability suggest that planetesimals can readily form beyond $100\,\textrm{au}$. We explore this circularisation by dynamical friction with a set of numerical simulations. We find that a planet that is scattered from the region close to Neptune onto an eccentric orbit has a 20-30\% chance of obtaining an orbit similar to that of Planet Nine after $4.6\, \textrm{Gyr}$. Our simulations also result in strong or partial clustering of the planetesimals; however, whether or not this clustering is observable depends on the location of the inner edge of the planetesimal belt. If the inner edge is located at $200\, \textrm{au}$ the degree of clustering amongst observable objects is significant.
\end{abstract}




\section{Introduction}

\citet{TrujilloSheppard2014} discovered an unexpected clustering in the argument of perihelion of minor planets with semimajor axis beyond $150\, \textrm{au}$ and perihelion beyond the orbit of Neptune, objects referred to as extreme trans-Neptunian objects (ETNOs). Subsequent orbital element analysis performed by \citet{BatyginBrown2016} showed that the orbits of these clustered ETNOs are physically aligned. The authors of both papers demonstrated that this clustering could be explained by the presence of a distant eccentric planet. 

\citet{TrujilloSheppard2014} demonstrated that a planet with mass between $2$ and $15\, \textrm{M}_{\oplus}$ and a semimajor axis beyond $200\, \textrm{au}$ could have created the clustering of ETNOs, and maintained it for billions of years. Stronger constraints were placed by \citet{BatyginBrown2016}, who showed that the clustering of longitude of perihelion and ascending node could be maintained by a planet of mass $\geq 10\, \textrm{M}_{\oplus}$ on a distant eccentric orbit that is inclined and anti-aligned with the ETNOs. They found in their simulations that orbits outside the parameter region bounded by semimajor axis $a \sim 400-1500\, \textrm{au}$ and eccentricity $e \sim 0.5-0.8$ were disfavoured. In this incarnation, the undetected planet has become known as "Planet Nine". This region was refined in a later paper by \citet{BrownBatygin2016}, who identified a region with $a \sim 380-980\, \textrm{au}$, $e \sim 0.1-0.8$, masses between $5-20\, \textrm{M}_{\oplus}$ and an inclination ($i$) of approximately $30^{\circ}$ relative to the ecliptic. Other authors \citep{Malhotra2016,Becker2017,MillhollandLaughlin2017} have studied the dynamical stability of the clustered ETNOs and used resonance considerations to constrain the orbital parameters of Planet Nine, which yield solutions within the region identified by \citet{BrownBatygin2016}. 

Various mechanisms have been proposed to explain the origin of Planet Nine, e.g. capture from another star during a close encounter in the Sun's birth cluster \citep{LiAdams2016,Mustill2016,Parker2017}, or \textit{in-situ} formation by slow coagulation within a distant ring of planetesimals \citep{KenyonBromley2015,KenyonBromley2016}. A third option is that Planet Nine is a scattered ice giant, originating in the outer giant-planet region. 
Planets with masses in the predicted regime for Planet Nine can be scattered onto large semimajor axis orbits by giant protoplanets growing in the ice-giant region that are clearing their respective orbital domains (e.g. \citealt{Thommes1999,LevisonMorbidelli2007}). Outward scattering can also occur during instabilities in mature systems (see e.g. \citealt{Tsiganis2005,Nesvorny2011}). However, unless such scattered planets somehow circularize their orbits, subsequent scattering near perihelion eventually leads to ejection. One potential mechanism for circularizing the orbit is through dynamical friction with an extended massive disc of gas  \citep{BromleyKenyon2016}. 

In this work we propose a new mechanism for circularizing the orbit of a scattered Planet Nine, namely through dynamical friction with a massive planetesimal belt beyond $100\, \textrm{au}$. Formation at these large distances is a result of FUV photoevaporation, which reduces gas mass and triggers planetesimal formation through the streaming instability. In recent simulations of planetesimal formation via the streaming instability performed by \citet{Carrera2017}, a massive ($60-130\, \textrm{M}_{\oplus}$) planetesimal belt forms beyond $100\, \textrm{au}$. We will refer to such an ultracold belt of planetesimals as a {\it cryobelt}. The amount of planetesimals formed interior to $100\, \textrm{au}$ is heavily dependent on the model and parameters that are being used, whilst the formation of a cryobelt by FUV photoevaporation is a robust result of the simulations and does not require particular fine tuning of any parameters (except for the assumption that FUV radiation indeed leads to efficient photoevaporation in the first place, as pointed out by \citealt{Ercolano2017}). Since this cryobelt forms early it should be present at the time of the giant planet formation, when Planet Nine is most likely to have been scattered out. An additional result from \citet{Carrera2017} is that the streaming instability together with FUV photoevaporation does not form enough planetesimals in the giant planet region to form the cores of giant planets, suggesting that the planetesimals taking part in the formation of giant planet cores form through some other process, e.g. particle pile-ups near ice lines (\citealt{Sirono2011a,Sirono2011b,IdaGuillot2016,DrazkowskaAlibert2017,SchoonenbergOrmel2017}). 

In this paper we investigate whether dynamical friction with a massive planetesimal belt beyond $100\, \textrm{au}$ can put a scattered ice giant on an orbit similar to that which is expected for the hypothetical Planet Nine. In Section 2 we briefly describe the numerical methods that have been used; in Section 3 we present results from the preliminary integrations; in Section 4 we present results from the final simulations; and in Section 5 we discuss and summarize the results. 

\section{Numerical methods}
In order to investigate the effect that a massive cryobelt has on the orbit of a scattered, eccentric ice giant, we have employed a series of direct \textit{N}-body simulations. The simulations were carried out using the \textsc{mercury} package \citep{Chambers1999}. We used the hybrid symplectic-Bulisch-Stoer algorithm throughout the paper with a tolerance parameter of $10^{-13}$, and set the timestep to be a twentieth of the shortest dynamical timescale, the orbital period of Uranus. Additional simulations with smaller tolerance parameters were executed in order to check that the outcome was not affected. 

A planet of mass $10\, \textrm{M}_{\oplus}$ was initiated on an eccentric orbit with a perihelion distance of $30\, \textrm{au}$, mimicking a recent scattering by Neptune. The initial inclination was set to be 5 degrees, since preliminary integrations showed that smaller initial inclinations resulted in systems with high eccentricities of Uranus and Neptune. 
Following the results from \citet{Carrera2017}, we introduce a massive cryobelt that has a surface density distribution following 1/$a$. We set the width of the cryobelt to be $500\, \textrm{au}$, and perform simulations for inner edges at $100$ and $200\, \textrm{au}$. The models of \citet{Carrera2017} produce cryobelts of masses down to $\gtrsim 60\, \textrm{M}_{\oplus}$. We use the lower mass limit since it results in a lower limit on the effect on the orbit of the scattered planet. We also perform a small set of simulations with a less massive cryobelt, to represent the scenario when mass is stripped from the solar system during its time in the birth cluster, or if planetesimal formation is less efficient than in our fiducial model. The cryobelt is represented by either $1000$ or $10000$ massive small bodies in \textsc{mercury} (see discussion in \ref{section:numberSmall}).

At an initial semimajor axis of $30\, \textrm{au}$, Planet Nine is unlikely to suffer any close encounters with Jupiter or Saturn, and so for the sake of computational time we chose to only include three big bodies in the simulations (Uranus, Neptune and Planet Nine). However, even though close encounters between Planet Nine and the gas giants might be unlikely, the gas giants could dominate the orbital evolution of some close-in cryobelt objects. As this is the part of the cryobelt that is most likely to be observable, we choose to use the same trick as \citet{BatyginBrown2016} and include the gravitational potential of the gas giants in a manner that does not demand a decreased time-step. We incorporated the acceleration due to the gas giants by increasing the radius of the Sun out to the orbit of Saturn ($R_* = a_S$) and adding a $J_2$ moment to its potential. The magnitude of the $J_2$ moment due to Jupiter and Saturn is \citep{Burns1976}
\begin{equation}
J_2 = \frac{1}{2}\sum_{i=1}^{i=2}\frac{m_ia_i^2}{M_*R_*^2}
\end{equation}
where $m_i$ is the mass of planet $i$ and $M_*$ is the mass of the Sun. 

The aim of the simulations was to investigate whether it is possible to circularize the orbit of Planet Nine enough to be within the region of parameters identified by \citet{BrownBatygin2016}, without considerably exciting the outer giant-planet region. We started off with a set of simulations with integration time $1\, \textrm{Gyr}$. Once the simulations were finished we identified those with orbital parameters within the region identified by \citet{BrownBatygin2016} and continued to run those for the age of the Solar system. We reject runs with an energy error of greater than $10^{-3}$; however, most of our runs have energy errors of order $10^{-5}$.
\section{Preliminary integrations}

\subsection{Sensitivity to number of small bodies}\label{section:numberSmall}
\begin{figure}
\centerline{\includegraphics[width = \columnwidth]{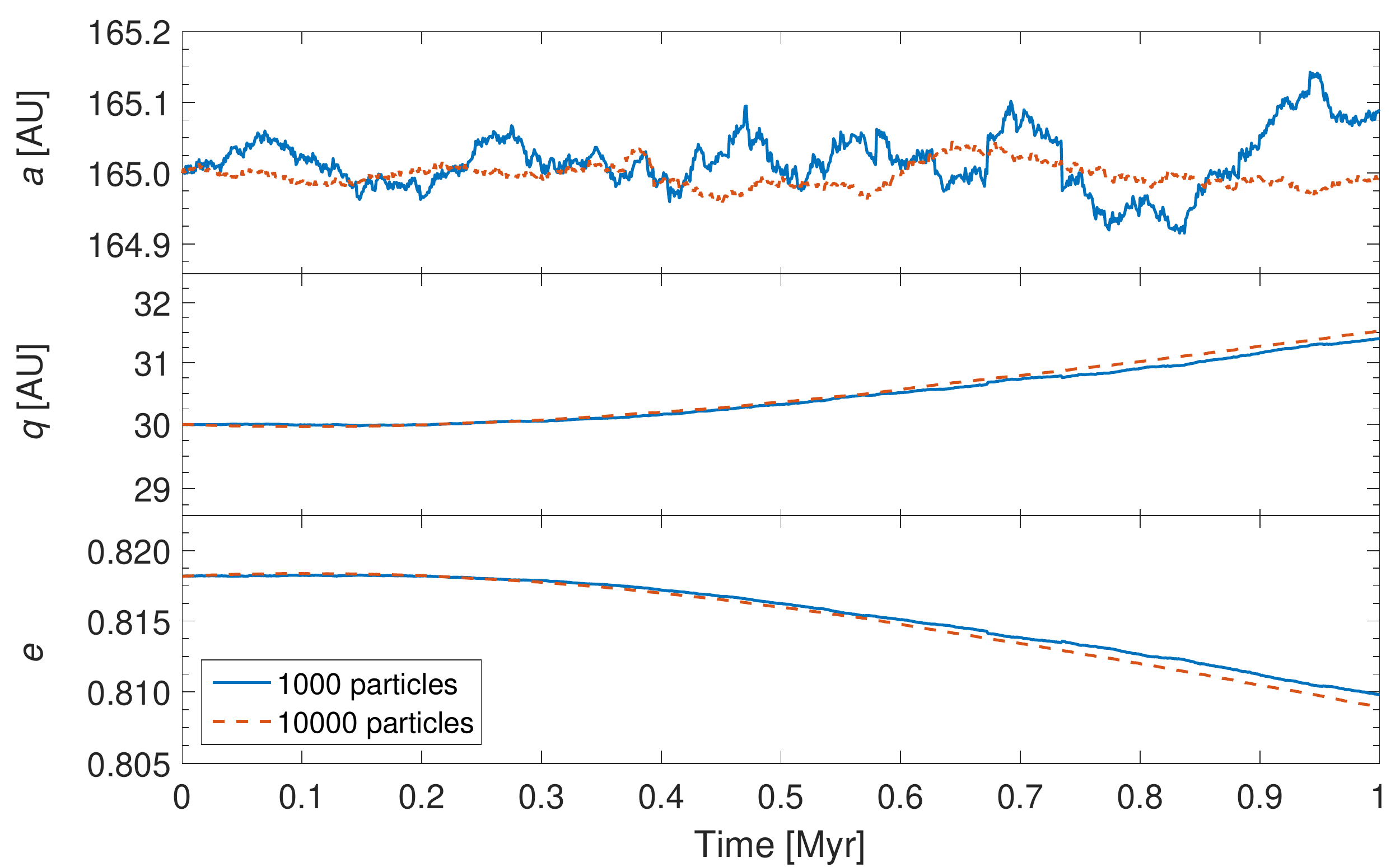}}
\caption{Sensitivity of the simulations to the number of small bodies. The figure shows the evolution of the semimajor axis (top), the perihelion distance  (middle) and the eccentricity (bottom) for a system containing a scattered ice giant and a massive cryobelt made up of either 1000 (blue) or 10000 (red) small bodies.}
\label{fig:1Pmanysmall}
\end{figure}

In order to test the sensitivity of the scattered planet's orbit to the number of small bodies, we integrated the orbit of the scattered planet in the presence of a cryobelt consisting of either 1000 or 10000 small bodies. The mass of the cryobelt was uniformly distributed between the small bodies. The system was integrated for 500 orbits, after which the evolution of the orbital elements was compared (see Fig.~\ref{fig:1Pmanysmall}). The stability of the orbit was improved with an increased number of small bodies, while the evolution of the eccentricity and perihelion distance diverged only slightly. As the study does not aim for a very high precision, and since the computational time can be greatly diminished with a smaller number of bodies, the small divergence was deemed negligible, and future simulations were performed with 1000 small bodies. 

The mass of each cryobelt object in the simulations is $60/1000 \, \textrm{M$_{\oplus}$} = 0.06\, \textrm{M$_{\oplus}$}$. For comparison, the mass of the Kuiper belt is estimated to be around $0.04 - 0.1\, \textrm{M}_{\oplus}$, and hundreds of thousands of KBOs are thought to exist. The most massive planetesimals formed by the streaming instability have a mass of around $10^{-4} - 10^{-5}\, \textrm{M}_{\oplus}$ (\citealt{Johansen2015,Simon2016}). In other words, the cryobelt objects used in our simulations are super-particles and represent many smaller objects. 

\subsection{Lack of excitation of the Kuiper belt}
Depending on the initial inclination of Planet Nine, it takes around $10-100\, \textrm{Myr}$ to lift the perihelion of the orbit out of the Kuiper belt. Before this occurs it has to pass through the Kuiper belt. The high velocity at the perihelion passing should make the scattering effect on the objects in the Kuiper belt fairly low, but in order to establish this we performed a small set of simulations including Uranus, Neptune, Planet Nine and a low-mass Kuiper belt.

\begin{figure}
\includegraphics[width = \columnwidth]{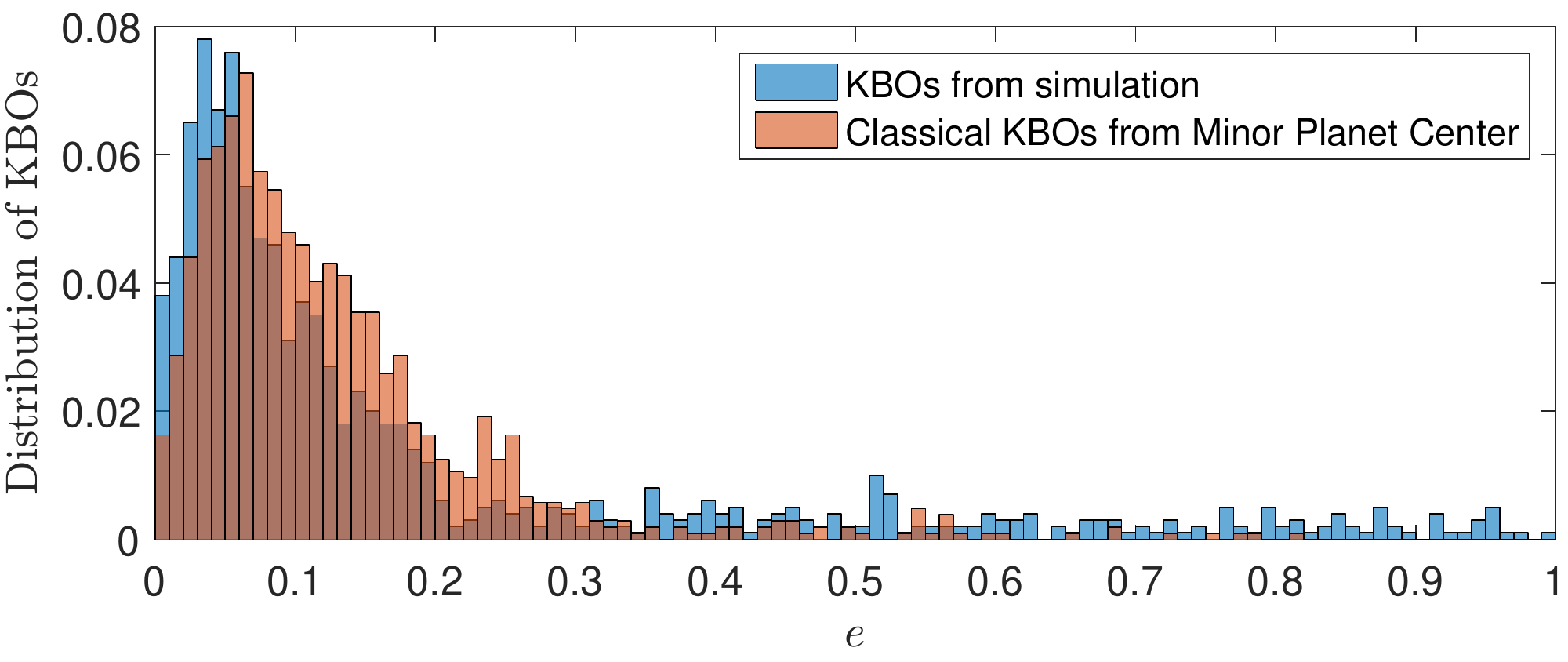}
\caption{Excitation of the Classical Kuiper Belt by an eccentric Planet 9 being scattered by the ice giants. The histogram shows the final eccentricities of objects from the simulation that were initially within the Kuiper belt (blue) and the eccentricities of known classical Kuiper belt objects (red). The excitation of the Kuiper belt due to Planet Nine results in a distribution of KBOs that well resembles that of the classical Kuiper belt. }
\label{fig:histogram}
\end{figure}

We performed 3 different simulations with 1000 KBOs that were given semimajor axes between $30$ and $55\, \textrm{au}$ and eccentricities and inclinations between 0.001 and 0.01 at the beginning of the integration. Planet Nine was given different initial semimajor axis in each simulation: the values tested were $400$, $600$ and $800\, \textrm{au}$. After $100\, \textrm{Myr}$ more than $50\%$ of the Kuiper belt objects had eccentricities below $0.1$, and over $75\%$ of them had eccentricities below $0.2$ ( see Figure \ref{fig:histogram}). This result is true for all simulations and did not vary with the initial semimajor axis of Planet Nine. We also checked the inclination excitation and found that over $75\%$ of the Kuiper belt objects had inclinations smaller than $5^{\circ}$, similar to the cold classical KBOs. As we have no cryobelt in these simulations Planet Nine's perihelion does not detach from the Kuiper Belt region, so this sets an upper limit on KBO excitation. Excitation could be avoided altogether if Planet Nine were scattered prior to the Kuiper Belt being emplaced, as may be the case if the classical Kuiper belt formed at the latest stages of protoplanetary disc evolution, as seen in the simulations of \citet{Carrera2017}.

\section{Results}
\begin{table}
\centering
\caption{Parameter choices for our simulations.}
\label{table:parameters}
\begin{tabular}{lcc}
\hline
\multicolumn{1}{c}{Parameter} & Value or Range                     & Units                \\ \hline
{\ul \textit{Planet Nine}}    & \multicolumn{1}{l}{}               & \multicolumn{1}{l}{} \\
Mass                          & 10                                 & $M_{\oplus}$          \\
Initial inclination           & 5                                  & deg              \\
Initial semimajor axis       & 400, 500, 600, 700, & au                   \\
                               & 800, 900, 1000       & \multicolumn{1}{l}{} \\
Initial perihelion            & 30                                 & au                   \\
{\ul \textit{Cryobelt}}       & \multicolumn{1}{l}{}               & \multicolumn{1}{l}{} \\
Mass                          & 60 (10,20,30)                                 & $M_{\oplus}$          \\
Number of particles           & 1000                               & \multicolumn{1}{l}{} \\
Inner edge                    & 100, 200                           & au                   \\ 
Outer edge					  & 600, 700						   & au					  \\
\hline
\end{tabular}
\end{table}
\subsection{Scattering of Planet Nine}
\begin{figure}
\centering
\includegraphics[width= 1\columnwidth]{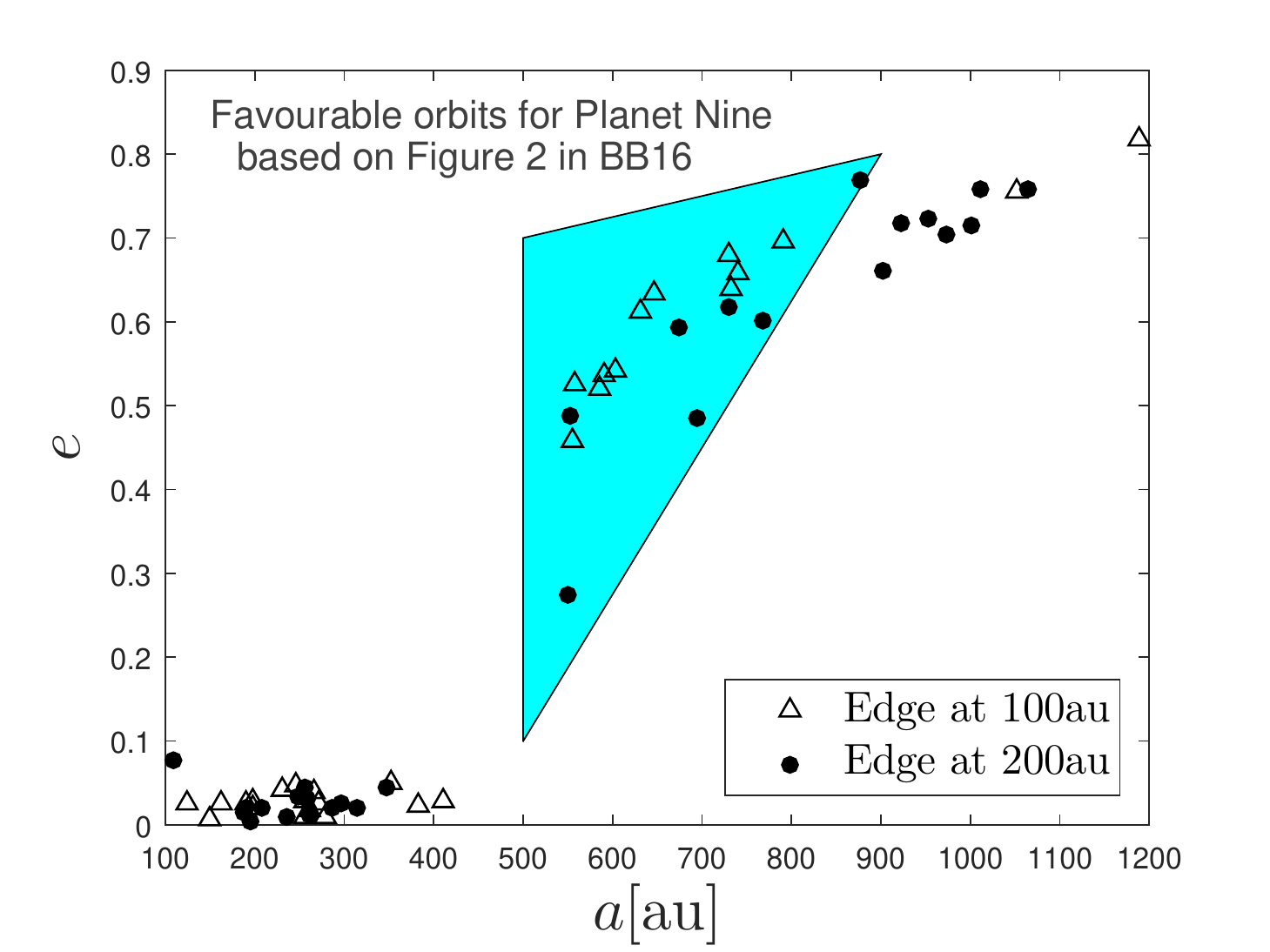}
\caption{The values of eccentricity and semimajor axis for Planet Nine at the end of each integration. The blue triangular area represents favorable orbits of Planet Nine and is based on Figure 2 from \citet{BrownBatygin2016}. Simulations which result in Planet Nine parameters within the blue region is considered successful. In this plot we show results from simulations with a cryobelt edge at $100\, \textrm{au}$ (triangular marker) and results from simulations with a cryobelt edge at $200\, \textrm{au}$ (circular marker). }
\label{fig:P9aVSaMustill}
\end{figure}

\begin{figure*}
\centering
\includegraphics[width= 2\columnwidth]{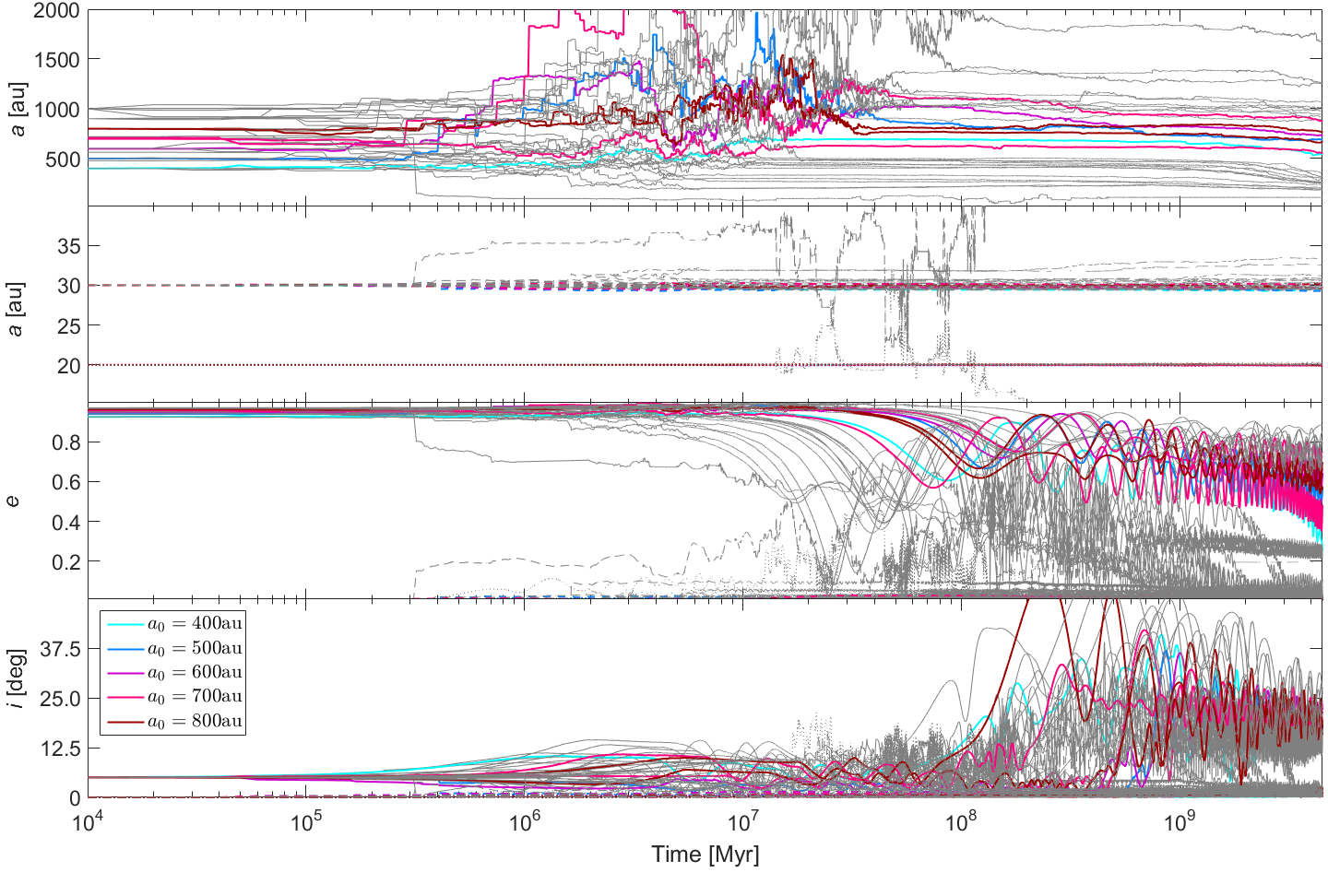}
\caption{The evolution of the semimajor axis for Planet Nine (top panel), the semimajor axis for Uranus and Neptune (second panel), the eccentricity for all three planets (third panel) and the inclinations for all three planets (bottom panel) from all simulations performed with an inner edge of the cryobelt at $200\, \textrm{au}$. Simulations in which Planet Nine ended up within the region identified by \citet{BrownBatygin2016} (the ones considered successful in Figure \ref{fig:P9aVSaMustill}) are colour-coded after their initial semimajor axis. Thin grey lines represent simulations where Planet Nine did not end up within this region. In the figure dotted lines represent Uranus, dashed lines Neptune and solid lines Planet Nine. Due to continual encounters between Neptune and Planet Nine happening around $1-10\, \textrm{Myr}$ after the simulations start, the final semimajor axis of Planet Nine varies considerably from the initial one. It is the outcome of these close encounters that decides whether or not Planet Nine will end up in the region identified by \citet{BrownBatygin2016}. } 
\label{fig:successful}
\end{figure*}

\begin{figure}
\centering
\includegraphics[width=1\columnwidth]{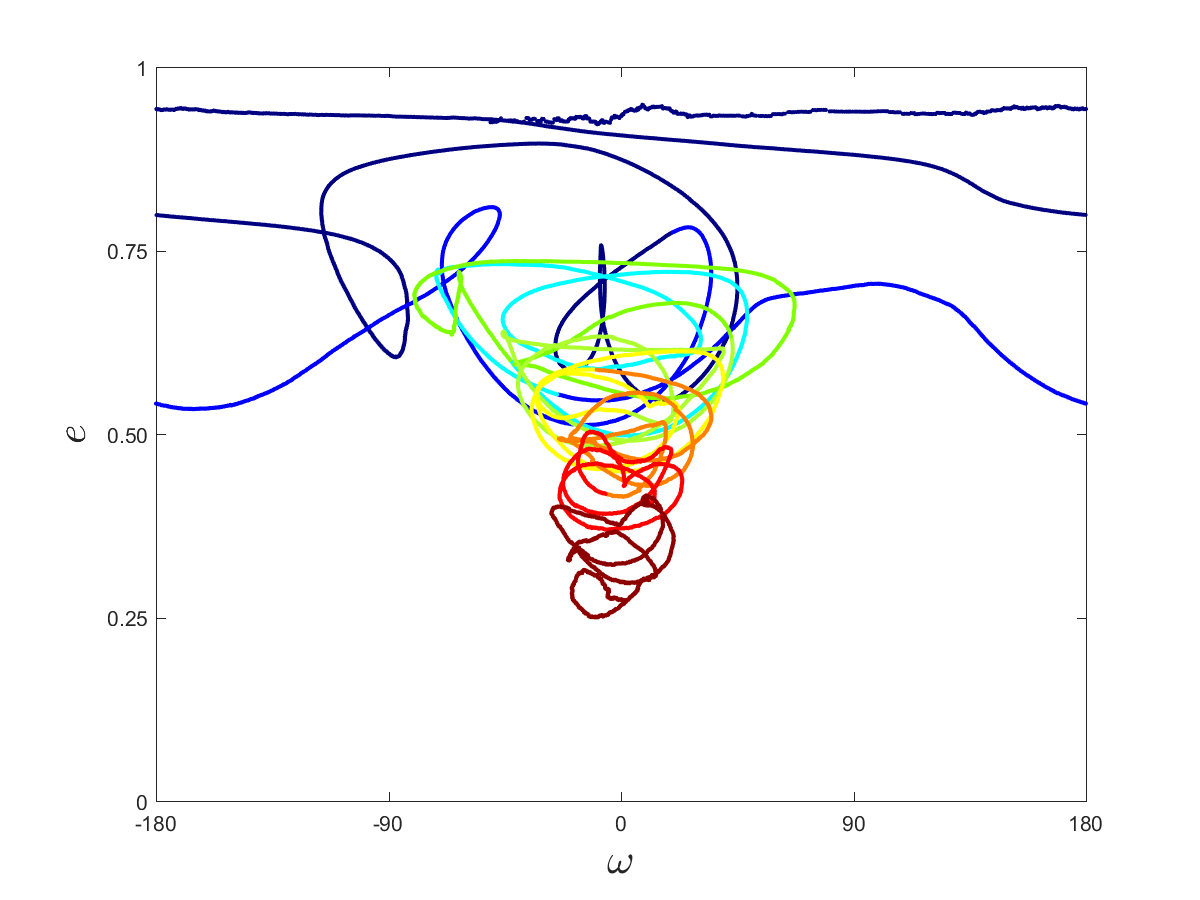}
\caption{The eccentricity versus argument of perihelion of Planet Nine as the two orbital elements evolve over time (blue to red). Planet Nine starts with high eccentricity at the top of the plot. As the eccentricity decreases, Planet Nine eventually gets trapped in a Kozai resonance with the more massive cryobelt. This perturbation give rise to oscillations about a constant value of the orbit's argument of perihelion, resulting in eccentricity being traded for inclination. In this particular simulation Planet Nine had an initial semimajor axis of $400\, \textrm{au}$. }
\label{fig:kozai}
\end{figure}

We started off our simulations with Planet Nine already on an eccentric orbit with perihelion at $30\, \textrm{au}$. In order to test how strong the initial scattering of Planet Nine must have been we performed a series of simulations where Planet Nine was initiated on an orbit with perihelion distance just outside Neptune but with varying semimajor axes; see Table \ref{table:parameters} for the different initial configuration of Planet Nine. We performed 35 simulations each for an inner edge of the cryobelt at either $100\, \textrm{au}$ or $200\, \textrm{au}$, each with a different random seed for the cryobelt particles.

In Figure \ref{fig:P9aVSaMustill} the final eccentricities and semimajor axes of Planet Nine from all performed simulations can be viewed. The blue area represents favourable orbits for Planet Nine and is based on Figure 2 from \citet{BrownBatygin2016}; the area is chosen in a similar manner as in \citet{Mustill2016}. Simulations which result in Planet Nine parameters within this blue area, while at the same time leaving Uranus and Neptune at orbital eccentricities below 0.05, are considered successful. Out of all simulations with an inner edge of the cryobelt at $100\, \textrm{au}$, approximately 30\% (11/35) where successful. For simulations with an inner edge of the cryobelt at $200\, \textrm{au}$ the success rate was 20\% (7/35). 

The evolution of the semimajor axis, eccentricity and inclination of the three ice giants is shown in Figure \ref{fig:successful} for simulations with an inner edge of the cryobelt at $200\, \textrm{au}$. Looking at the successful simulations we obtained a maximum final orbital eccentricity of 0.02 for Uranus, and 0.03 for Neptune. The preferred inclination from \citet{BrownBatygin2016} was $30^{\circ}$, however they find in their calculations that an inclination of $20^{\circ}$, which is what we obtain in our simulations, works approximately as well.

From Figure \ref{fig:successful} we find that the simulations that failed to end up within the desired parameter region either circularized down towards zero, or suffered too much scattering by Neptune before becoming detached from the giant planet region. All simulations suffered continual close encounters by Neptune at integration time between $1-10\, \textrm{Myr}$. We find in our simulations that the initial choice of semimajor axis for Planet Nine does not affect the result. This is at least true for our particular choice of initial perihelion distance for Planet Nine. We obtained successful simulations for initial semimajor axes between $400$ and $800\, \textrm{au}$. 

Pure dynamical friction results in inclination damping \citep{Popolo1999}; however, we find in our simulations that the inclination of Planet Nine increases (see lower panel of Figure \ref{fig:successful}). This hints to some other mechanism operating in the system apart from dynamical friction. In our simulations Planet Nine starts with a very high eccentricity and a low inclination. With time the eccentricity of Planet Nine will decrease due to dynamical friction, and once it reaches a sufficiently low value Planet Nine becomes trapped in a Kozai resonance (\citealt{Kozai1962,Lidov1962}) with the more massive cryobelt. The perturbation on Planet Nine's orbit then gives rise to oscillations about a constant value of the orbit's argument of perihelion. In Figure \ref{fig:kozai} we show a typical example of the trajectory followed by Planet Nine, in the phase space of eccentricity versus argument of perihelion. Initially Planet Nine lies in the region of high-eccentricity circulating trajectories at the top of the figure. Here, eccentricity remains high and barely fluctuating, and so the orbital inclination remains low. As eccentricity is damped, it enters the Kozai resonance ``from above" and begins the oscillatory exchange of eccentricity and inclination. This entry into the resonance from above contrasts with the situation usually considered for the Kozai effect, where eccentricity starts low and inclination high. Entry into the resonance is a robust feature of our simulations. This mechanism thus explain the oscillatory behavior of Planet Nine's orbital inclination and eccentricity seen in the two lower panels of Figure \ref{fig:successful}; as well as the overall increase in inclination of Planet Nine's orbit to the values favored by \citet{BrownBatygin2016}.

The main goal of this paper was to investigate whether it was possible to circularize a scattered ice giant enough to be within the parameter region predicted for Planet Nine via dynamical friction with a cryobelt. The result from this first subsection shows that this is indeed possible.

\subsection{Simulations with a lower-mass cryobelt}
During the early life of the solar system, when the sun was still in its birth cluster, stellar flybys were likely common. During such flybys there is a chance that mass from far out in the solar system gets stripped. As the cryobelt is located several hundred au away from the sun, it is not impossible for some of this mass to be stripped from the solar system before the orbit of Planet Nine has been put in place. If too much mass is being stripped too early, it might not be possible to circularize the orbit of Planet Nine. Additionally, planetesimal formation might not be quite as prolific as in the models of \citet{Carrera2017}.

To check this we performed a small set of simulations with a lower cryobelt mass. The limiting case would be when the mass of the cryobelt is the same as the mass of Planet Nine. We performed 5 simulations of this and found that Planet Nine was ejected in $4/5$ simulations. In the last simulation Planet Nine ended up fairly close to the region indicated in Figure \ref{fig:P9aVSaMustill}; however, the final orbital inclination was close to zero and the semimajor axis of Neptune got shifted a few au. When a cryobelt with twice the mass of Planet Nine was used instead, the success rate increased substantially. Out of 5 simulations Planet Nine became trapped in Kozai resonances and ended up inside or near the desired region twice, and was only ejected once. Finally we performed the same set of simulations for a cryobelt three times more massive than Planet Nine, and although none of the 5 ended up inside the desired region, this time no simulation resulted in ejection of Planet Nine.

All of the simulations performed with a lower mass cryobelt resulted in an orbital inclination close to zero for Planet Nine. This is not the case for the previous simulations with a $60\, \textrm{M}_{\oplus}$ cryobelt. In other words, if the mass of the cryobelt is large then Planet Nine naturally gets excited to a higher inclination. However, if the mass of the cryobelt is small then Planet Nine must acquire a $20-30$ degree inclination through the initial scattering event. Our preliminary integrations show that such an initial inclination does not affect the orbits of either Uranus or Neptune; in fact such systems are generally more stable than systems where Planet Nine is given an initial inclination of $5$ degrees. 

\subsection{Clustering and observability of the cryobelt particles}
\begin{figure}
\includegraphics[width= \columnwidth]{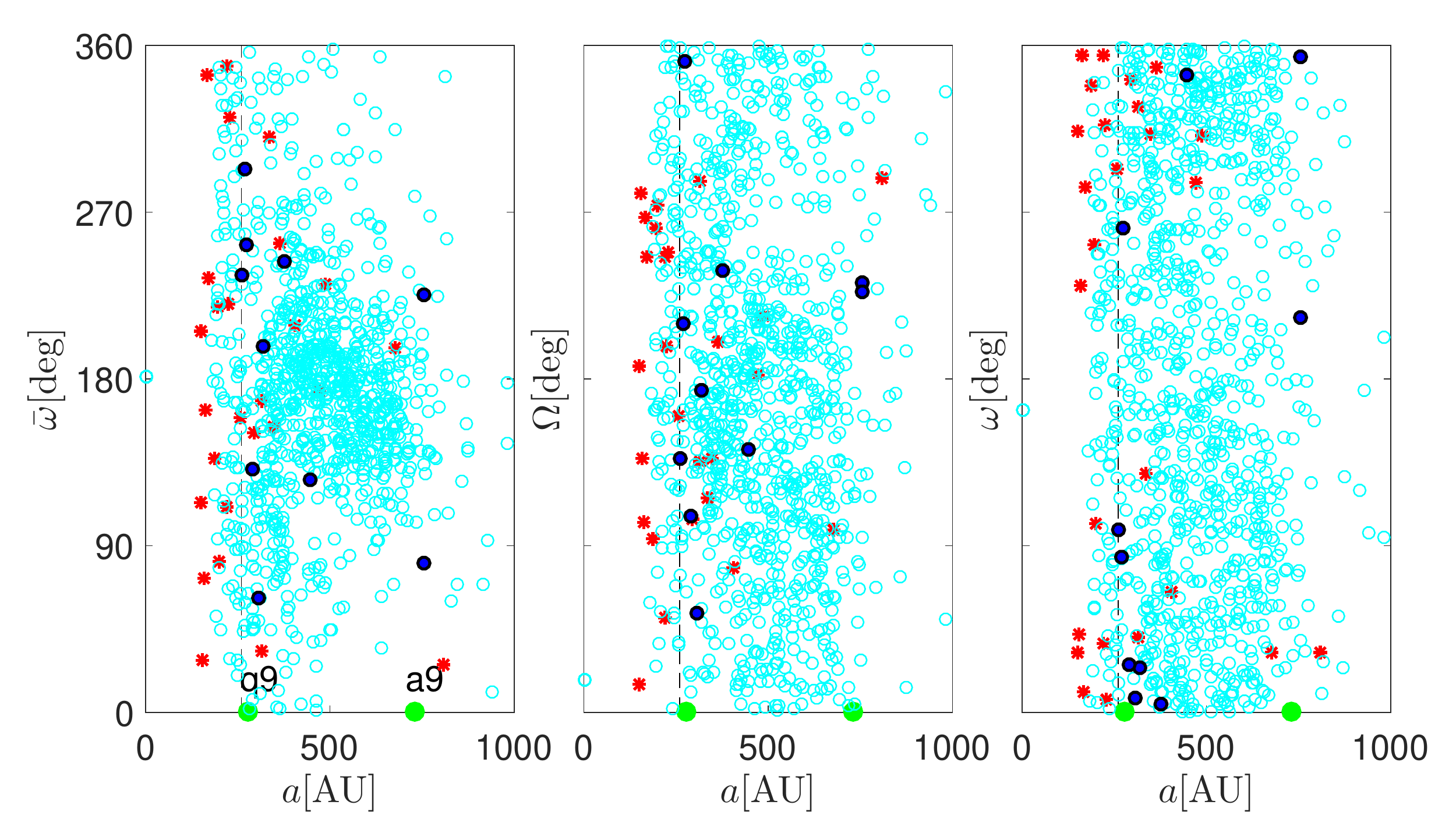}
\includegraphics[width= \columnwidth]{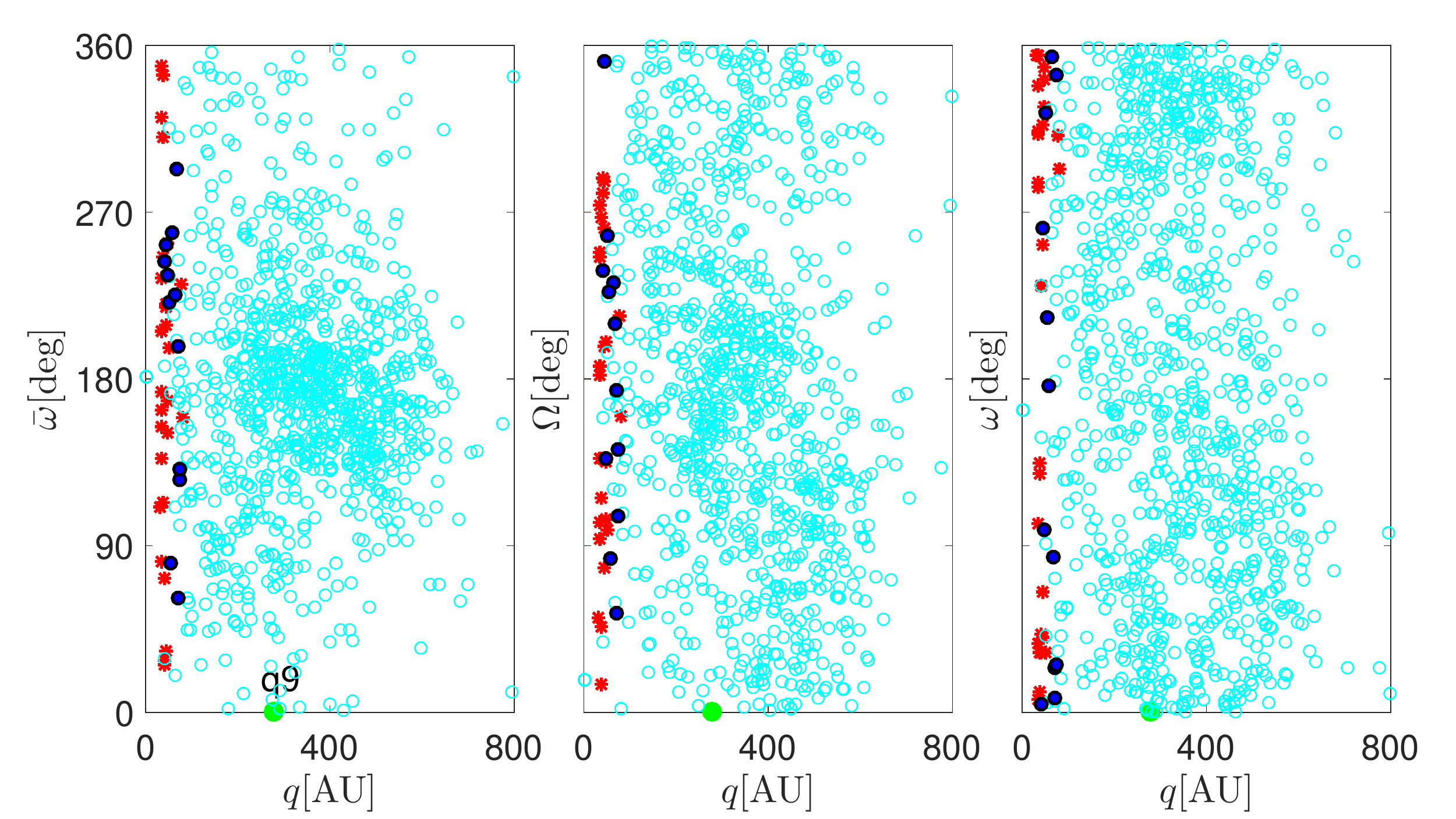}
\caption{The plots are from one of the successful simulations with an inner edge of the cryobelt at $200\, \textrm{au}$ and show the longitude of perihelion (left), longitude of ascending node (middle) and argument of perihelion (right) of all surviving objects in the cryobelt as a function of semimajor axis (top) and perihelion (bottom). Cryobelt objects with perihelion inside 2012VP113 and semimajor axis beyond the dashed line are marked with dark-blue circles; these are in theory observable and dominated by Planet Nine. The positions of the cryobelt objects in the figure are relative to Planet Nine, which is positioned at 0 degrees in the plot of longitude of perihelion and longitude of ascending node. Data of known TNOs has been added to the plot; these are marked with red filled dots and are positioned relative to Planet Nine. In the left and middle plots cryobelt objects are clustering at $180^{\circ}$ away from Planet Nine, and this clustering is also visible in the right plots. It is clear from the bottom plots that clustering increases with increasing perihelion of the cryobelt objects.}
\label{fig:orbital_a}
\end{figure}

\begin{figure}
\includegraphics[width= \columnwidth]{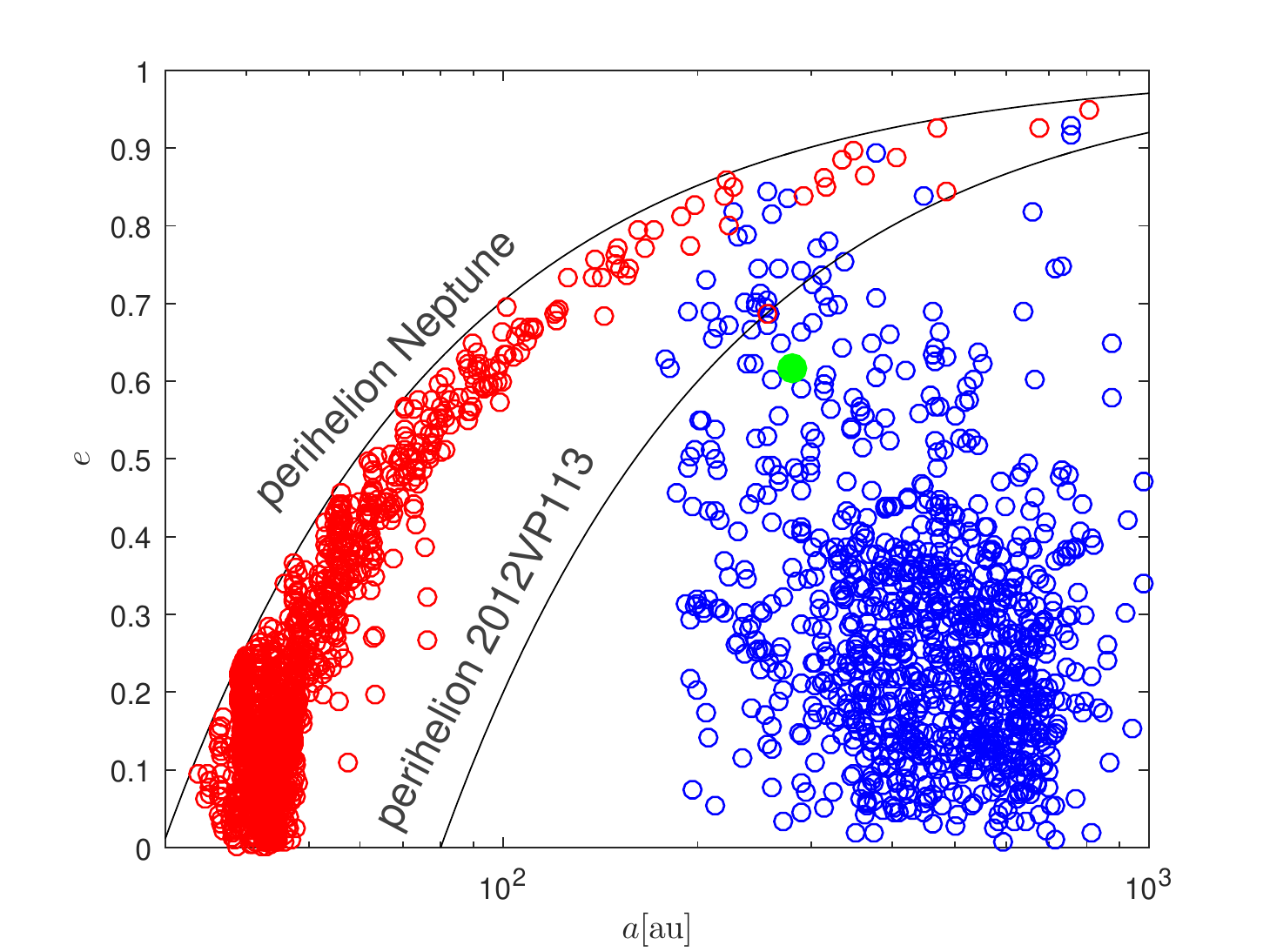}
\caption{Eccentricity versus semimajor axis for cryobelt objects from one of the successful simulations (blue) and known TNOs from the Minor Planet Center (red). The perihelion of Neptune and of 2012VP113 (the most distant observed TNO) are marked in the plot, and the position of Planet Nine has been added as a green dot. Most Cryobelt objects remain far out in the solar system; however, a small fraction mix with the scattered disc and should be observable. }
\label{fig:eccentricity}
\end{figure}

\begin{figure}
\centering
\includegraphics[width = .8\columnwidth]{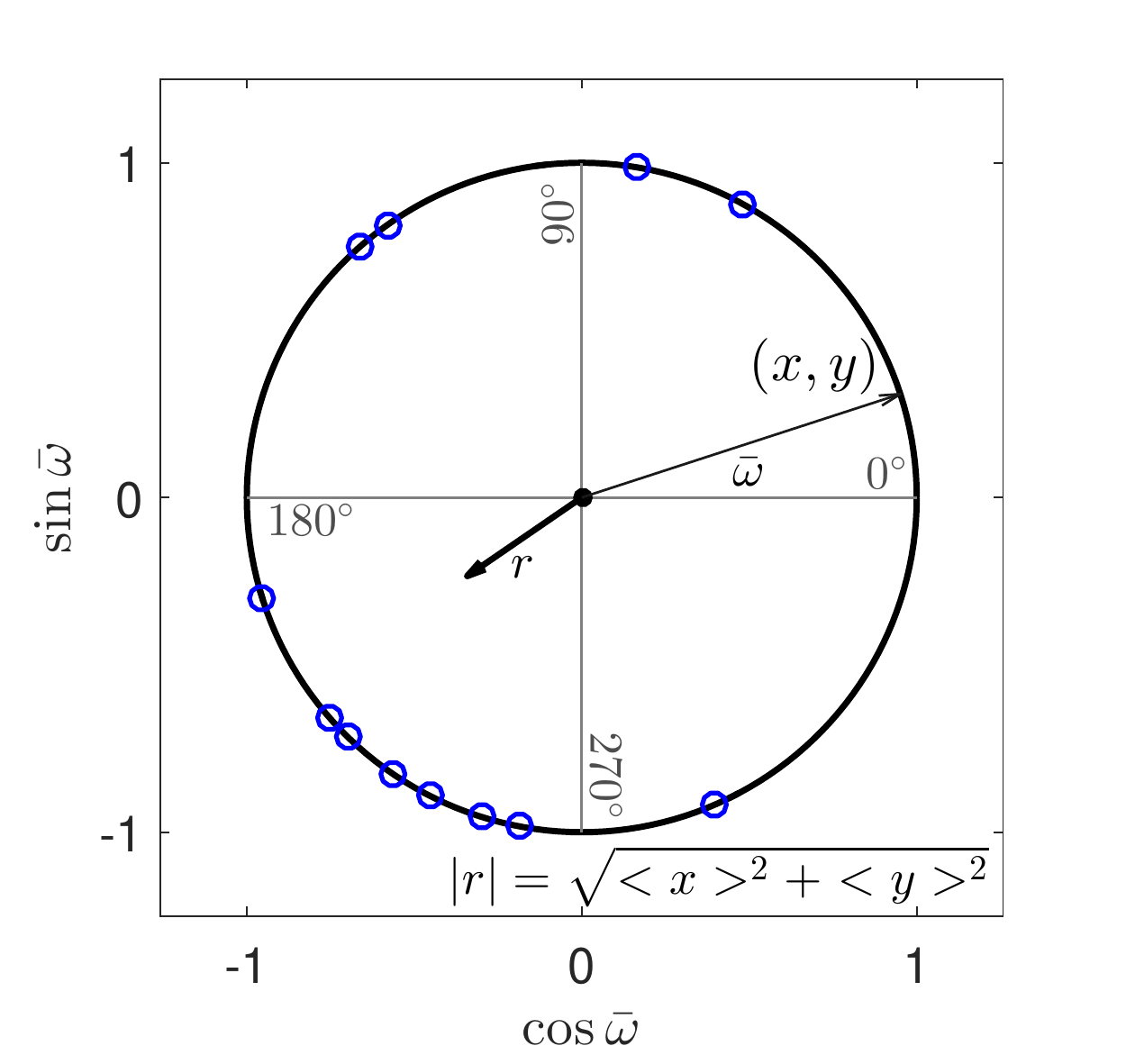}
\caption{Example where we show how the amount of clustering can be calculated for the dark-blue circles in Figure \ref{fig:orbital_a}. If the longitude of perihelion angles for all objects are converted to Cartesian coordinates; then the length of a vector $\bm{r}$ pointing towards the mean of these $x$ and $y$ coordinates tell us how clustered the data points are. If the data points are homogeneously distributed over $360$ degrees the length of this vector will be 0, while equal angles results in a vector of length unity. In this example the length of the vector is $0.42$. }
\label{fig:vectorCalculation}
\end{figure}

\begin{figure}
\includegraphics[width=\columnwidth]{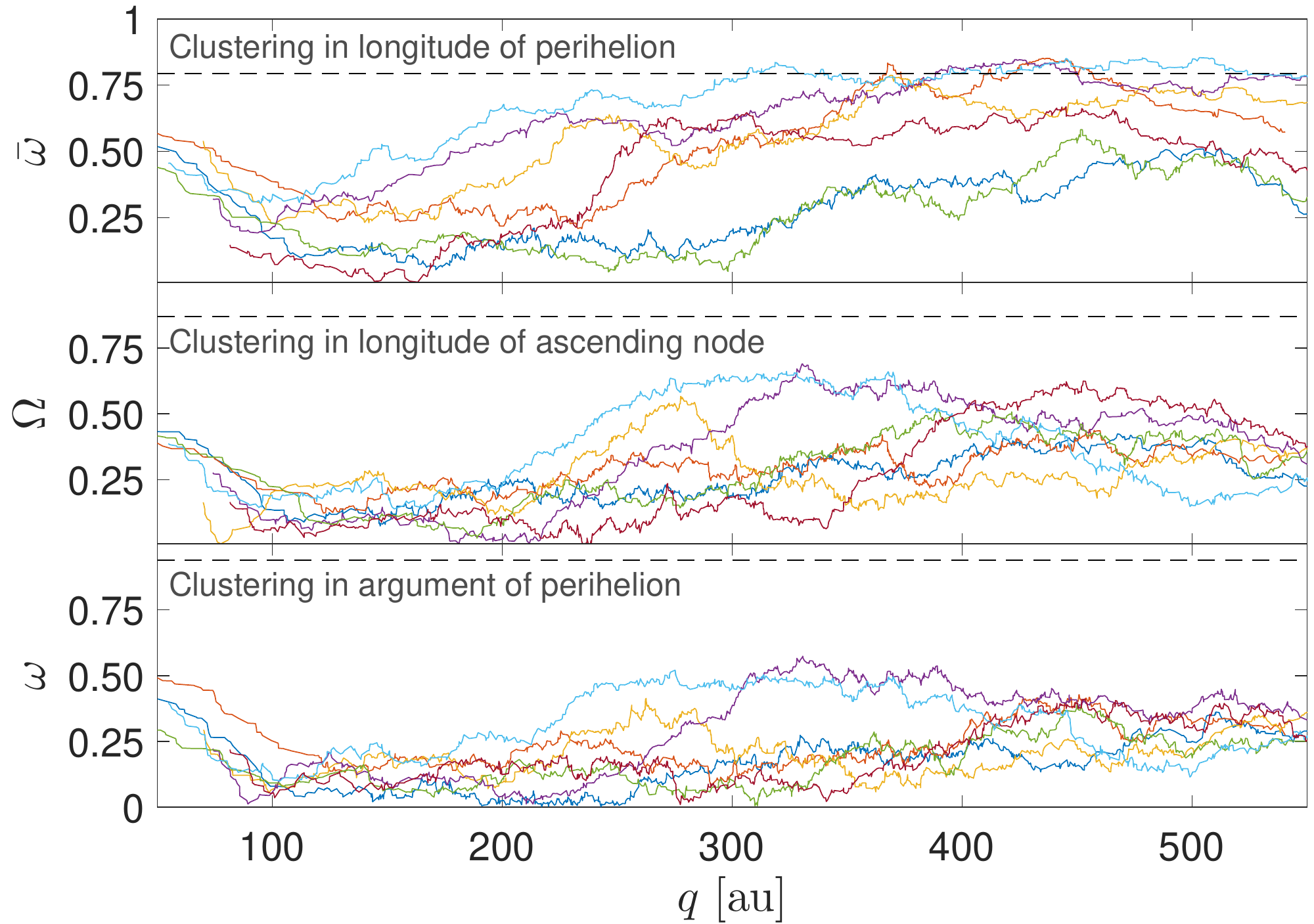}
\caption{The plot shows the degree of clustering in the longitude of perihelion (top), longitude of ascending node (middle) and the argument or perihelion (bottom) as a function of perihelion distance for all of the successful simulations. For comparison we also calculate $|\bm{r}|$ for the six objects with $a > 250\, \textrm{au}$ from Figure 1 in \citet{BatyginBrown2016} and add that as a dashed line. Most of the clustering in our data occurs around $300-500\, \textrm{au}$. In the observable part of the solar system the degree of clustering is lower but still significant, with values of $|\bm{r}|$ reaching up to around $0.5$. See the result section and Figure \ref{fig:vectorCalculation} for a description of how the degree of clustering is calculated.}
\label{fig:vectorAngles}
\end{figure}

The observations that led to the Planet Nine hypothesis are the proposed clustering of ETNO's, which many authors have demonstrated can be created and maintained by a distant eccentric planet (e.g. \citealt{TrujilloSheppard2014,BatyginBrown2016}). As our simulations include a cryobelt positioned well beyond Neptune's orbit, we can not only test if this clustering occurs but also if it could persist for the age of the Solar System. 

In Figure \ref{fig:orbital_a} we show the longitude of perihelion, longitude of ascending node and argument of perihelion for the cryobelt particles at the end of one of the successful integrations. The inner part of the cryobelt is dominated by the giant planets and experience little or no clustering. Objects with semimajor axes outside $270\, \textrm{au}$, however, experience strong or partial clustering in all three orbital angles. This result is true for all successful simulations, although the edge beyond which clustering starts to appear can shift a few tens of au. The location of this edge depends on the inner edge of the cryobelt and the position of Planet Nine. When the inner edge of the cryobelt is moved from $100\, \textrm{au}$ to $200\, \textrm{au}$, the part of the cryobelt that is affected by the giant planets is greatly diminished. 

As we have postulated a large population of distant bodies in the cryobelt, we need to ensure that this population is compatible with observed TNO populations. We found that, when setting the inner edge of the cryobelt to $100\, \textrm{au}$, the simulations yielded large numbers of planetesimals with a semimajor axis of $\sim 100\, \textrm{au}$ and perihelia smaller than 2012VP113, a population which is not observed. We therefore also ran simulations where the inner edge of the cryobelt was set to $200\, \textrm{au}$. These latter simulations resulted in fewer observable cryobelt objects and better resembles the known population of ETNOs. In Figure \ref{fig:eccentricity} we show the eccentricities and semimajor axes for all surviving cryobelt objects from one of the successful simulations with an inner edge of the cryobelt at $200\, \textrm{au}$. Interactions with Planet Nine leave some of the cryobelt objects on excited orbits with eccentricities as high as $0.9$. A significant number of these objects obtain perihelion distances less than 2012VP113 and mix in with the scattered disc.

One interesting feature that we see in all of our simulations, is that the clustering of the cryobelt objects mostly occurs beyond the perihelion of the most distant observed ETNO, 2012VP113 at $q=80\, \textrm{au}$. We investigate this further by calculating the degree of clustering amongst the cryobelt objects as a function of their perihelion distances. To do so we use a method similar to circular spectral analysis (see e.g. \citealt{Lutz1985}). The first step is to sort all cryobelt objects by perihelion distance. Then we loop through all cryobelt objects in this order, and for each object convert the angles of the orbits of the 100 closest lying objects to Cartesian coordinates. The length of a vector $\bm{r}$ pointing towards the mean of these coordinates tell us how clustered the data is. Only if $|\bm{r}| = 0$ are the cryobelt objects distributed homogeneously over the unit circle. If, on the other hand, $|\bm{r}| = 1$ then this implies that all of the cryobelt objects are perfectly aligned. See Figure \ref{fig:vectorCalculation} for a pictorial description of how the calculation is performed.

In Figure \ref{fig:vectorAngles} we show the degree of clustering in longitude of perihelion, longitude of ascending node and argument of perihelion as a function of perihelion distance for all successful simulations performed with an inner edge of the cryobelt at $200\, \textrm{au}$. The degree of clustering is largest around $q=300-500\, \textrm{au}$, close to the location of Planet Nine's perihelion. Cryobelt objects with perihelion distances smaller than that of 2012VP113 experience significant clustering with values of $|\mathbf{r}|$ up to around $0.5$. The degree of clustering is decreasing with perihelion distance up to $q \approx 100 - 150\, \textrm{au}$, when this trend reverses. In contrast, simulations with an inner edge of the cryobelt at $100\, \textrm{au}$ results in values of $|\mathbf{r}|$ that is below $0.25$ for objects with perihelion distance less than $150\, \textrm{au}$. Such a low degree of clustering is hard to detect unless the sample is big, and in some simulations it does not exist at all. In other words, the clustering that we obtain in simulations with an inner edge of the cryobelt at $100\, \textrm{au}$ is not observable. The simulations with an inner edge of the cryobelt at $200\, \textrm{au}$ however result in clustering that is much more likely to be observable. The low-perihelion cryobelt objects obtain high eccentricities and mix in with the scattered disc; see Figure \ref{fig:eccentricity}. Our simulations could thus explain the clustering in argument of perihelion seen amongst ETNO's by \citet{TrujilloSheppard2014} to a certain degree.  

\section{Conclusions}
In this paper we have shown that dynamical friction with an ultracold belt of planetesimals, which we have dubbed a cryobelt, can put a scattered, eccentric planet on an orbit similar to that inferred for Planet Nine. In other words, Planet Nine attains its orbit, with a raised perihelion and moderate inclination, as a natural outcome of planetary instability among the giant planets of the solar system, likely a direct consequence of the formation of multiple protoplanets in the ice giant region (e.g. \citealt{Nesvorny2011,Izidoro2015}). This means that we do not need to invoke a relatively unlikely capture event (e.g. \citealt{Mustill2016,Parker2017,Perets2012}) or a mechanism of in-situ formation of a planetary mass object at several hundred au from the sun \citep{KenyonBromley2015,KenyonBromley2016}. We considered planets of mass $10\, \textrm{M}_{\oplus}$, initiated on an eccentric orbit with perihelion close to Neptune. We performed a set of simulations to map which scattering events could result in orbital parameters within the regime identified by \citet{BrownBatygin2016}; however, in our simulations we found that the initial choice of semimajor axis for Planet Nine does not seem to affect the result. This is at least true for our particular choice of initial perihelion distance, where Planet Nine is coupled dynamically to Neptune after the initial scattering event.

Continual interactions between Planet Nine and the cryobelt particles lead to strong or partial clustering in the orbital angles of the cryobelt objects, similar to that seen in observations. This clustering occurs for all successful simulations, and can thus not be used to further refine the initial conditions. Cryobelt objects which obtain perihelion distances less than $80\, \textrm{au}$ experience a significant degree of clustering in simulations with an inner edge of the cryobelt at $200\, \textrm{au}$. The degree of clustering is substantially lower for simulations performed with an inner edge of the cryobelt at $100\, \textrm{au}$. \citet{TrujilloSheppard2014} find that all known objects with semi-major axis greater than $150\, \textrm{au}$ and perihelion greater than that of Neptune experience this clustering. Our simulations could explain some of the clustering but the degree of clustering is far from 100\%.   

Simulations with an inner edge of the cryobelt at $100\, \textrm{au}$ generally produce more mass within the observable region of the solar system than seen in observations. This problem is remedied by moving the inner edge of the cryobelt out to $200\, \textrm{au}$. Most of the initial mass of the cryobelt remains at the end of the simulations; around 4-5\% enters the giant planet region or is ejected from the system, and a few percentage mix in with the scattered disc. This would imply that there is substantial mass sitting beyond the visible region of the solar system. Except for collisional grinding, which as we state further down is ineffective, stellar flybys is the only event that could possibly work to reduce this mass. If the mass of the cryobelt had decreased substantially by such perturbations during the last $4\, \textrm{Gyr}$, then the orbit of Planet Nine would have already been in place (see Figure \ref{fig:successful}) and thus our result would likely not have changed much. However, such late flybys are rare and it is not certain how much mass that could be lost in this way. In the early life of the solar system stellar flybys were more common; however, if too much mass is lost too early it might not be possible to circularize Planet Nine's orbit. To check this we performed a small set of simulations with a lower cryobelt mass. We found that our model still yields good analogues for Planet Nine for cryobelt masses down to $20\, \textrm{M}_{\oplus}$ .

Objects in the cryobelt form \textit{in-situ} via the streaming instability, and the size distribution is therefore likely to resemble that of the cold classical Kuiper Belt (\citealt{Johansen2015,Simon2016}). The slope in the size distribution of the classical Kuiper belt is steeper than that of the hot population of the Kuiper belt and the scattered disc \citep{Fraser2014}. The slope gets much steeper for larger objects, and due to this the cold classical Kuiper belt lacks very large objects. If this is also true for the cryobelt, then cryobelt objects beyond $100\, \textrm{au}$ will be very hard to observe. Observed ETNOs are generally in the dwarf-planet mass range, likely a reflection of a formation region much closer to the Sun and subsequent outwards scattering during Neptune's migration. Cryobelt objects, on the other hand, should maintain the birth size distribution of planetesimals, since the growth rate by pebble accretion and mutual collisions is so low where these objects form.  

If cryobelts exist, could such planetesimal populations be observable around other stars? Planetesimal belts are inferred through the presence of debris discs, so in order to detect a planetesimal belt they must be able to create and maintain a debris disc. Since debris discs are formed through collisions between planetesimals and the resulting collisional cascade, there must be enough collisions between the planetesimals in the belt in order to produce a debris disc. The amount of collisions is determined by the collisional lifetime of the planetesimals, $t_c$. We calculate the collisional lifetime of the largest planetesimal in our cryobelt in the same manner as \citet{Wyatt2007}. The planetesimal density, maximum planetesimal radius and size distribution of planetesimals is obtained from \citet{Johansen2015}. The dispersal threshold, $Q_D^*$, was taken to be $200\, \textrm{J kg}^{-1}$. The calculation results in collisional lifetimes that are longer than the age of the solar system in all parts of the cryobelt. This means that no debris disc will be created, and the cryobelt will thus not be detectable by direct observations. 

During the circularization process approximately 2 Earth masses of planetesimals are ejected from the solar system by Planet Nine. These ejected planetesimals could possibly have maintained bound orbits in the inner Oort cloud (e.g. \citealt{Hills1981}), if we had included galactic shear in our simulations. If that would have been the case, a fraction of them could return to the inner solar system as comets which can be observed. Then there are the planetesimals that mix in with the scattered disc. Since they form in situ far out in the disc, their chemical composition will differ from most other scattered disc objects. In the future, if observations get precise enough, measurements of chemical composition of scattered disc objects might be able to tell if part of the high-eccentricity population formed very far out in the protoplanetary disc. This part of the population could then be identified as interlopers from the cryobelt. 

Whereas we have used a cryobelt for circularizing a scattered Planet Nine, \citet{BromleyKenyon2016} investigate whether is could be circularized by a gas disc. They considered transition discs and performed simulations with different sizes of the inner cavity, which in our simulations is equivalent to changing the inner edge of the cryobelt. They found that a large central cavity is essential to end up within the wanted region of parameters, and were able to obtain good results for inner edges of both $100\, \textrm{au}$ and $200\, \textrm{au}$, which is what we have used in our simulations. The demand put on the gas disc in order to land within the desired region of parameters is that it is either low-mass and long-lived, or high-mass and short-lived. In order to circularize the orbit of Planet Nine sufficiently within a few Myr, which is the expected time for disc dissipation, they find that the initial disc mass would have had to be about $0.1\, \textrm{M}_{\odot}$ or more. The advantage of our model is that the lower mass limit on the cryobelt is a robust outcome from simulations of planetesimal formation via the streaming instability, and that this cryobelt is expected to persist for far longer than any gas disc. 

\section*{Acknowledgements}
The authors wish to thank Konstantin Batygin and the anonymous referee for helpful comments that led to an improved manuscript. The authors also wish to thank the anonymous referee for suggesting an investigation of the role of the Kozai effect on the dynamics.
The authors are supported by the project grant ``IMPACT" from the Knut and Alice Wallenberg Foundation. AJ was further supported the Knut and Alice Wallenberg Foundation grants 2012.0150 and 2014.0048, the Swedish Research Council (grant 2014-5775) and the European Research Council (ERC Consolidator Grant 724687-PLANETESYS).
This research has made use of data and/or services provided by the International Astronomical Union's Minor Planet Center.





\begin{thebibliography}{99}


\bibitem[Batygin \& Brown(2016)]{BatyginBrown2016} Batygin, K., \& Brown, M.~E.\ 2016, \aj, 151, 22 
\bibitem[Becker et al.(2017)]{Becker2017} Becker, J., Adams, F., Khain, T., Hamilton, S., \& Gerdes, D.\ 2017, arXiv:1706.06609 
\bibitem[Bromley \& Kenyon(2016)]{BromleyKenyon2016} Bromley, B.~C., \& Kenyon, S.~J.\ 2016, \apj, 826, 64 
\bibitem[Brown \& Batygin(2016)]{BrownBatygin2016} Brown, M.~E., \& Batygin, K.\ 2016, \apjl, 824, L23
\bibitem[Burns(1976)]{Burns1976} Burns, J.~A.\ 1976, American Journal of Physics, 44, 944 
\bibitem[Carrera et al.(2017)]{Carrera2017} Carrera, D., Gorti, U., Johansen, A., \& Davies, M.~B.\ 2017, \apj, 839, 16
\bibitem[Chambers(1999)]{Chambers1999} Chambers, J.~E.\ 1999, \mnras, 304, 793 
\bibitem[Del Popolo et al.(1999)]{Popolo1999} Del Popolo, A., Spedicato, E., \& Gambera, M.\ 1999, \aap, 350, 685
\bibitem[Dr\k{a}\.{z}kowska \& Alibert(2017)]{DrazkowskaAlibert2017} Dr\k{a}\.{z}kowska, J., \& Alibert, Y.\ 2017, arXiv:1710.00009 
\bibitem[Ercolano et al.(2017)]{Ercolano2017} Ercolano, B., Jennings, J., Rosotti, G., \& Birnstiel, T.\ 2017, \mnras, 472, 4117 
\bibitem[Fraser et al.(2014)]{Fraser2014} Fraser, W.~C., Brown, M.~E., Morbidelli, A., Parker, A., \& Batygin, K.\ 2014, \apj, 782, 100 
\bibitem[Hills(1981)]{Hills1981} Hills, J.~G.\ 1981, \aj, 86, 1730 
\bibitem[Ida \& Guillot(2016)]{IdaGuillot2016} Ida, S., \& Guillot, T.\ 2016, \aap, 596, L3 
\bibitem[Izidoro et al.(2015)]{Izidoro2015} Izidoro, A., Morbidelli, A., Raymond, S.~N., Hersant, F., \& Pierens, A.\ 2015, \aap, 582, A99 
\bibitem[Johansen et al.(2015)]{Johansen2015} Johansen, A., Mac Low, M.-M., Lacerda, P., \& Bizzarro, M.\ 2015, Science Advances, 1, 1500109
\bibitem[Kenyon \& Bromley(2015)]{KenyonBromley2015} Kenyon, S.~J., \& Bromley, B.~C.\ 2015, \apj, 806, 42
\bibitem[Kenyon \& Bromley(2016)]{KenyonBromley2016} Kenyon, S.~J., \& Bromley, B.~C.\ 2016, \apj, 825, 33
\bibitem[Kozai(1962)]{Kozai1962} Kozai, Y.\ 1962, \aj, 67, 591 
\bibitem[Levison \& Morbidelli(2007)]{LevisonMorbidelli2007} Levison, H.~F., \& Morbidelli, A.\ 2007, \icarus, 189, 196
\bibitem[Li \& Adams(2016)]{LiAdams2016} Li, G., \& Adams, F.~C.\ 2016, \apjl, 823, L3 
\bibitem[Lidov(1962)]{Lidov1962} Lidov, M.~L.\ 1962, \planss, 9, 719 
\bibitem[Lutz(1985)]{Lutz1985} Lutz, T.~M.\ 1985, \nat, 317, 404 
\bibitem[Malhotra et al.(2016)]{Malhotra2016} Malhotra, R., Volk, K., \& Wang, X.\ 2016, \apjl, 824, L22
\bibitem[Millholland \& Laughlin(2017)]{MillhollandLaughlin2017} Millholland, S., \& Laughlin, G.\ 2017, \aj, 153, 91 
\bibitem[Mustill et al.(2016)]{Mustill2016} Mustill, A.~J., Raymond, S.~N., \& Davies, M.~B.\ 2016, \mnras, 460, L109 
\bibitem[Nesvorn{\'y}(2011)]{Nesvorny2011} Nesvorn{\'y}, D.\ 2011, \apjl, 742, L22 
\bibitem[Parker et al.(2017)]{Parker2017} Parker, R.~J., Lichtenberg, T., \& Quanz, S.~P.\ 2017, \mnras, 472, L75
\bibitem[Perets \& Kouwenhoven(2012)]{Perets2012} Perets, H.~B., \& Kouwenhoven, M.~B.~N.\ 2012, \apj, 750, 83 
\bibitem[Schoonenberg \& Ormel(2017)]{SchoonenbergOrmel2017} Schoonenberg, D., \& Ormel, C.~W.\ 2017, \aap, 602, A21 
\bibitem[Simon et al.(2016)]{Simon2016} Simon, J.~B., Armitage, P.~J., Li, R., \& Youdin, A.~N.\ 2016, \apj, 822, 55 
\bibitem[Sirono(2011a)]{Sirono2011a} Sirono, S.-i.\ 2011, \apjl, 733, L41 
\bibitem[Sirono(2011b)]{Sirono2011b} Sirono, S.-i.\ 2011, \apj, 735, 131 
\bibitem[Thommes et al.(1999)]{Thommes1999} Thommes, E.~W., Duncan, M.~J., \& Levison, H.~F.\ 1999, \nat, 402, 635
\bibitem[Trujillo \& Sheppard(2014)]{TrujilloSheppard2014} Trujillo, C.~A., \& Sheppard, S.~S.\ 2014, \nat, 507, 471 
\bibitem[Tsiganis et al.(2005)]{Tsiganis2005} Tsiganis, K., Gomes, R., Morbidelli, A., \& Levison, H.~F.\ 2005, \nat, 435, 459 
\bibitem[Wyatt et al.(2007)]{Wyatt2007} Wyatt, M.~C., Smith, R., Greaves, J.~S., et al.\ 2007a, \apj, 658, 569 


\end{thebibliography}








\bsp	
\label{lastpage}
\end{document}